\documentclass[journal]{IEEEtran}
\usepackage{bm}
\usepackage{xcolor}
\usepackage{caption}
\usepackage{graphicx}
\usepackage{epstopdf}
\usepackage{algorithm}
\usepackage{newtxtext}
\usepackage{algorithmicx}
\usepackage[fleqn]{amsmath}
\usepackage[varg]{newtxmath}
\newcommand{\cA}{{\mathcal A}}
\newcommand{\cB}{{\mathcal B}}
\newcommand{\cD}{{\mathcal D}}
\newcommand{\cF}{{\mathcal F}}
\newcommand{\cO}{{\mathcal O}}
\newcommand{\cU}{{\mathcal U}}
\newcommand{\MA}{\boldsymbol A}
\newcommand{\MB}{\boldsymbol B}
\newcommand{\MI}{\boldsymbol I}

\newcommand{\MP}{\boldsymbol P}
\newcommand{\MD}{\boldsymbol D}
\newcommand{\MO}{\boldsymbol O}
\newcommand{\MQ}{\boldsymbol Q}

\newcommand{\vone}{\boldsymbol 1}

\newcommand{\vzero}{\boldsymbol 0}
\newcommand{\vth}{\boldsymbol \theta}

\DeclareMathOperator*{\argmax}{argmax}

\begin{document}
\title{Markovian Arrival Process Parameter Estimation of Quasi-birth-death Queueing Systems with Utilization Data}
\author{
Chen~Li,~\IEEEmembership{Senior Member,~IEEE,}
Junjun~Zheng,~\IEEEmembership{Member,~IEEE,}
Hiroyuki~Okamura,~\IEEEmembership{Member,~IEEE,}
and~Tadashi~Dohi,~\IEEEmembership{Member,~IEEE}
\thanks{C. Li is with the D3 Center, The University of Osaka, Japan. e-mail: li.chen.d3c@osaka-u.ac.jp}
\thanks{J. Zheng is with the Department of Information Engineering, Graduate School of Advanced Science and Engineering, Hiroshima University, Japan. e-mail: jzhenghiroshima-u.ac.jp}
\thanks{H. Okamura is with the Department of Information Engineering, Graduate School of Advanced Science and Engineering, Hiroshima University, Japan. e-mail: okamu@hiroshima-u.ac.jp}
\thanks{T. Dohi is with the Department of Information Engineering, Graduate School of Advanced Science and Engineering, Hiroshima University, Japan. e-mail: dohi@hiroshima-u.ac.jp}
}
\markboth{Journal of IEEE trans.,~Vol.~11, No.~4, December~2023}
{Li \MakeLowercase{\textit{et al.}}: MAP Parameter Estimation of QBD Queueing Systems with Utilization Data}
\maketitle
\begin{abstract}
Parameter estimation for queueing systems is commonly performed using inter-arrival times, waiting times, or queue-length observations. However, such detailed observations are often unavailable in practical computer systems, where utilization data, such as CPU utilization, is much easier to collect. Utilization data provides only the fraction of time during which the system is busy within each monitoring interval, while the exact arrivals, services, phase transitions, and system states in unobservable periods remain hidden. This paper proposes an expectation-maximization (EM) algorithm for estimating the parameters of Markovian arrival process (MAP)-driven quasi-birth-death (QBD) queueing systems from utilization data. The proposed method formulates the underlying queueing dynamics as a QBD process and derives the expected sufficient statistics for sojourn times, phase transitions, arrivals, and services over both observable and unobservable intervals. These expectations are then used to iteratively update the MAP and service parameters under the maximum likelihood framework. In addition, Akaike's information criterion is introduced to select the appropriate number of MAP phases and mitigate overfitting. The proposed framework enables MAP-based queueing parameter estimation from incomplete utilization observations and provides a practical modeling approach for systems where detailed event-level measurements are difficult to obtain.
\end{abstract}
\IEEEpeerreviewmaketitle

\section*{Nomenclature}
\addcontentsline{toc}{section}{Nomenclature}
{\it Acronyms and Abbreviations}
\begin{IEEEdescription}[\IEEEusemathlabelsep\IEEEsetlabelwidth{MMPP}]
\item[MAP] Markovian arrival process.
\item[MMPP] Markov modulated Poisson process.
\item[BMAP] Batch Markovian arrival process.
\item[CTMC] Continuous-time Markov chain.
\item[MLE] Maximum likelihood estimation.
\item[EM] Expectation maximization.
\item[PH] Phase type.
\item[HMM] Hidden Markov chain.
\item[HPP] Homogeneous Poisson process.
\item[NHPP] Non-homogeneous Poisson process.
\item[FCFS] First come first served.
\item[LLF] Log-likelihood function.
\item[QBD] Quasi-birth-death process.
\item[AIC] Akaike’s information criterion.
\end{IEEEdescription}

\vspace{0.5cm}
{\it Notations}
\begin{IEEEdescription}[\IEEEusemathlabelsep\IEEEsetlabelwidth{$\vzero_{1\times m(K+1)}$}]
\item[$K$] Capacity of the $MAP/M/1/K$ queueing system.
\item[$\MQ_{MAP}$] Infinitesimal generator matrix of MAP.
\item[$\MD_0$] Infinitesimal generator matrix in the case of no arrivals.
\item[$\MD_1$] Transition rate matrix when an arrival occurs.
\item[$N(t)$] Cumulative number of arrivals during the time interval $[0,t)$.
\item[$J(t)$] Phase process of the underlying CTMC at time $t$.
\item[$m$] Maximum number of phases of MAP.
\item[$q_{i,j}(i \neq j)$] Transition rate from phase $i$ to phase $j$.
\item[$\lambda_{i,j}$] Arrival rate from phase $i$ to phase $j$ of MAP.
\item[$\bm{\pi}_s$] Stationary probability vector of MAP.
\item[$\tilde{\lambda}$] Fundamental arrival rate.
\item[$\MP_k(t)$] Matrix whose element indicates that the number of arrivals is $k$ by time $t$.
\item[$\bm{\pi}$] Initial probability vector for phases.
\item[$\lambda_i$] Arrival rate of phase $i$ of MMPP.
\item[$\cD$] Observable data.
\item[$u_n$] $n$-th utilization sample of the observable period.
\item[$\cU$] Unobservable data.
\item[$\text{E}_{\cU}$] Expectation operator for the unobservable data $\cU$.
\item[$t_u$] Length of unobservable time interval.
\item[$t_o$] Length of observable time interval.
\item[$B_{t_o}^{(n)}$] Busy time for the $n$-th observable time interval.
\item[$I_{t_o}^{(n)}$] Idle time for the $n$-th observable time interval.
\item[$B_i$] An indicator random variable for the event that the phase is $i$ at the initial time $t=0$.
\item[$Z_i^{[n,l]}$] Cumulative sojourn time that the number of jobs is $l$ in phase $i$ during the $n$-th unobservable time interval.
\item[$N_{i,j}^{[n,l]}$] Number of phase transitions from phase $i$ to phase $j$ that the number of jobs is $l$ during the $n$-th unobservable time interval.
\item[$A_{i,j}^{[n,l]}$] Number of arrivals leading to phase transitions from phase $i$ to phase $j$ that the number of jobs is $l$ during the $n$-th unobservable time interval.
\item[$S_{i,j}^{[n,l]}$] Number of services leading to phase transitions from phase $i$ to phase $j$ that the number of jobs is $l$ during the $n$-th unobservable time interval.
\item[${\tilde Z}_i^{[n,l]}$] Cumulative sojourn time that the number of jobs is $l$ in phase $i$ during the $n$-th observable time interval.
\item[${\tilde N}_{i,j}^{[n,l]}$] Number of phase transitions from phase $i$ to phase $j$ that the number of jobs is $l$ during the $n$-th observable time interval.
\item[${\tilde A}_{i,j}^{[n,l]}$] Number of arrivals leading to phase transitions from phase $i$ to phase $j$ that the number of jobs is $l$ during the $n$-th observable time interval.
\item[${\tilde S}_{i,j}^{[n,l]}$] Number of services leading to phase transitions from phase $i$ to phase $j$ that the number of jobs is $l$ during the $n$-th observable time interval.
\item[$\vth$] A set of parameters.
\item[$\mu_{i,j}$] Service rate from phase $i$ to phase $j$ of MAP.
\item[$T$] Length of monitoring time interval.
\item[$s_n$] Cumulative monitoring time of $n$-th utilization sample.
\item[$s_n^-$] Left limit of $n$-th utilization sample $s_n$.
\item[$s_n^+$] Right limit of $n$-th utilization sample $s_n$.
\item[$\cF_n$] Forward events.
\item[$\cB_n$] Backward events.
\item[$P(A)$] Probability of any indicator random variable $A$.
\item[$\bm{f}(n)$] Probability vector for forward events at the beginning of $n$-th unobservable time interval.
\item[$\bm{b}(n)$] Probability vector for backward events at the beginning of $n$-th unobservable time interval.
\item[$\tilde{\bm{f}}(n)$] Probability vector for forward events at the beginning of $n$-th observable time interval.
\item[$\tilde{\bm{b}}(n)$] Probability vector for backward events at the beginning of $n$-th observable time interval.
\item[$\tilde{\bm{f}^{\prime}}(n)$] Probability vector of the system state change between busy and idle in the observable time interval for forward events.
\item[$\tilde{\bm{b}^{\prime}}(n)$] Probability vector of the system state change between busy and idle in the observable time interval for backward events.
\item[$\MQ_{QBD}$] Infinitesimal generator matrix of a QBD process queueing system.
\item[$\MI$] Identity matrix.
\item[$\MO$] $m \times m$ zero matrix.
\item[$\MQ_{00}$] Infinitesimal generator matrix when the system is idle.
\item[$\MQ_{01}$] Infinitesimal generator matrix when the system state changes from idle to busy.
\item[$\MQ_{10}$] Infinitesimal generator matrix when the system state changes from busy to idle.
\item[$\MQ_{11}$] Infinitesimal generator matrix when the system state changes from busy to busy.
\item[{$[\cdot]_{(l,i)}$}] $i$-th element of the $l$-th block in the block vector $[\cdot]$.
\item[{$[\cdot]_i$}] $i$-th element of the vector $[\cdot]$.
\item[{$[\cdot]_{i,j}$}] $(i,j)$-th element of the matrix $[\cdot]$.
\item[$\vzero_{1\times m(K+1)}$] $1 \times m(K+1)$ zero vector.
\item[$\vzero_{m(K+1)\times 1}$] $m(K+1)\times 1$ zero vector.
\item[$\MI$] $m\times m$ identity matrix.
\end{IEEEdescription}

\section{Introduction}
\label{sec:intro}
\IEEEPARstart{P}{arameter} estimation of queueing systems is a long-standing topic, which has been extensively studied in past decades \cite{Basawa:NRLQ:1981,Basawa:QRM:1992,Basawa:QS:1992,Mitchell:PE:2003}. Typically, a queueing system can be symbolized as $A/B/S/K$, where $A$ represents the arrival distribution, $B$ represents the service distribution, $S$ is the number of service nodes, and $K$ is the capacity of the queueing system. For example, a queueing model where the arrival process follows a Poisson distribution, the service time follows an exponential distribution, the number of service nodes is 1, and the queue length is $K$ can be represented as $M/M/1/K$. For a Poisson process, the renewal process requires a sequence of independent and identically distributed non-negative random variables formed between inter-arrival times. 

A Markovian arrival process (MAP) is a generalization of the Poisson process with dependence between the inter-arrival times and non-exponential inter-arrival time distribution \cite{Ushio:IJMMS:2009}. MAP was first proposed by Neuts \cite{Neuts:AP:1979} and has been widely used to analyze mathematically stochastic behaviors such as reliability, teletraffic, and performance evaluation \cite{Lucantoni:AIAP:1990,kim:CAOR:2010}. MAP is defined as a specific continuous-time Markov chain (CTMC) to represent the arrivals of a queueing system. A MAP consists of {\it phase} and {\it level} processes with corresponding discrete state spaces, respectively. The {\it phase} process is represented by a CTMC, which reflects the dynamic process of the internal state. The {\it level} process represents the number of events, which can be a counting process such as a Poisson process. MAP usually has two main properties. (i) MAP is the most versatile stochastic counting process, and it contains many arrival processes, such as the Poisson process, Markov-modulated Poisson process (MMPP) \cite{Fischer:PE:1993}, and batch MAP (BMAP) \cite{Lucantoni:SM:1991}. In addition, MAP is known to be dense \cite{Asmussen:JAP:1993} for any stochastic point process, which can approximate complex stochastic counting processes. (ii) MAP can represent the time correlation in arrival streams. The commonly used arrival processes, such as the homogeneous Poisson process (HPP), often assume that the inter-arrivals are independent. However, this assumption is unrealistic in practice since the inter-arrival times are usually correlated, such as the long-range dependence problem in communication networks \cite{Leland:IEEETrans:1994}. In actual Ethernet traffic, long-range dependence problems are caused by self-similar processes. Therefore, MAP-based queueing system modeling has become significant.

Generally, the parameters (e.g., arrival and service rates) can be estimated by observing the behavior of a queueing system during a specified time interval. For example, all the above literature collected inter-arrival times to estimate the parameters of the arrival stream. To date, moment-based and likelihood-based approaches are the two most widely used approaches to parameter estimation for MAP. The moment-based approach determines the MAP parameters to fit the theoretical moments to empirical moments from observed data \cite{Heffes:JSAC:1986,Yoshihara:TS:2001}. The likelihood-based approach aims to find the MAP parameters from the empirical data when their likelihood is maximum \cite{Klemm:PE:2003}. 

However, collecting and recording inter-arrival times is costly and unrealistic in real-world situations. Additional data that record the available information of queueing systems (e.g., waiting times and queue lengths) are easier to obtain that are used for parameter estimation \cite{Basawa:QR:1996,Basawa:SPL:2008}. However, such data are difficult to collect in a computer system from observation of a queue. Utilization data is defined as the ratio of the time interval during which the system is busy to a fixed monitoring time. In a computer system, utilization data such as CPU utilization is one of the most commonly used statistics to monitor CPU behavior during task execution. Most operating systems have the function to calculate CPU utilization by default. Unfortunately, utilization data belongs to incomplete data that only can be collected during observable time intervals, while the information in the unobservable periods is missing. Although utilization data is easy to obtain, exact inter-arrivals and waiting times are not known by observation. To the best of our knowledge, no studies have been conducted to estimate MAP parameters from utilization data due to the limitations of traditional parameter estimation approaches.

An expectation-maximization (EM) algorithm is an iterative way to compute maximum likelihood estimation (MLE) from incomplete data \cite{Okamura:ITON:2009, Dempster:JRSSB:1977, Wu:AS:1983}. Therefore, an EM algorithm is used to estimate the MAP parameters from utilization data. To further generalize the proposed approach, this study seeks to estimate the parameters of a MAP quasi-birth-death (QBD) queueing system. The main contributions of this study are summarized below:

\begin{itemize}
\item {\bf Data collection of utilization data:} unlike traditional observable data, the arrival process is unknown from utilization data. Additionally, utilization data contains unobservable and observable intervals, and only utilization with the observable interval can be collected.

\item {\bf Generalization of a queueing system via MAP and QBD:} MAP is a generalization of the Poisson process to represent more complex bursts and associated traffic streams. QBD is a generalization of the birth-death process and is a powerful tool for modeling many stochastic phenomena.

\item {\bf An EM algorithm-based parameter estimation approach:} unlike the traditional moment-based and MLE-based approaches, a nontrivial EM algorithm can compute the MLE for utilization data containing unobservable intervals.
\end{itemize}


\section{Related Works}
\label{sec:related_works}
A temporal point process is a stochastic process used to model a time series of binary events that occur at random intervals in continuous time \cite{Daley:ETAM:2003}. A continuous-time stochastic process can be modeled by a CTMC with a discrete state space.CTMC models have applications in many fields, such as traffic modeling \cite{Akar:SAC:1998, Klemm:TOOLS:2002}, earthquake prediction \cite{Kim:ICOML:2018}, and performance evaluation \cite{Dharmaraja:CC:2003, Bolch:CC:2006}.

\subsection{Parameter Estimation of MAP}
MAP is one of the most flexible stochastic processes and is defined as a specific CTMC. Since MAP can approximate any point process, it is often used to model general arrival and service processes in a queueing system \cite{Bruneo:Comput:2013, Zheng:Rel:2017, Klimenok:CCIS:2017, Vygovskaya:CCIS:2018}. Bruneo et al. \cite{Bruneo:Comput:2013} proposed a virtualized system with a regenerative policy to evaluate the performance of the arrivals following the Poisson process and MMPP. Experiments show that the proposed workload-based policy is superior to the time-based strategy. Zheng et al. \cite{Zheng:Rel:2017} considered the workload-based and time-based policies for arrival streams following MAP. They also analyzed the loss probability of the transactions and the average response time of the estimated parameters based on Markovian arrivals. Klimenok et al. \cite{Klimenok:CCIS:2017} estimated the parameters of a queueing system in which the arrival stream and service time followed MAP and exponential distribution, respectively. They assumed that the system has two servers and that the buffer size is infinite. Vygovskaya et al. \cite{Vygovskaya:CCIS:2018} estimated the parameters of the arrival process with a simple variant of MAP (i.e., MMPP) as the waiting times. Their queueing system can be represented as $MMPP/M/2$. Note that the collected time series data are observable.

\subsection{Parameter Estimation for Observable Data}
Generally, moment-based and likelihood-based estimations are the two major approaches to fit the observed data and MAP parameter estimation. The moment-based estimation approach can reduce computational costs compared with the likelihood-based estimation. 

In moment estimation, the parameters of MAP can be estimated by matching the observed data between the sample moments and the population moments \cite{Hazelton:IESC:2011}. Heffes et al. \cite{Heffes:IEEECommun:1986} provided an explicit formula to estimate the parameters of the two-state MMPP (i.e., MMPP(2)) from empirical moments by observing the number of arrivals. Anderson et al. \cite{Andersen:IEEECommun:1998} estimated the superposition parameters of the two-state MAP (i.e., MAP(2)) based on moments by collecting the number of arrivals of the switched Poisson process.
Yoshihara et al. \cite{Yoshihara:TS:2001} fitted the superposition of MMPP(2) using a moment estimation approach for self-similar traffic. Mitchell et al. developed an approximate model to estimate the parameters of the $G/G/1/N$ queueing system by observing inter-arrival times. Additionally, Telek et al. \cite{Telek:PE:2007} considered a model from the moment inter-arrival time distribution to estimate the parameters of MAP.

The likelihood-based estimation is used to find the optimal MAP parameters by maximizing the likelihood function from a series of observations \cite{Hendry:PUP:2007, Chambers:CRCP:2012}. Although MLE is a well-known method for parameter estimation of a queueing system, it has significant limitations for estimating the parameters of MAP. MAP consists of many parameters, which require large matrix operations. Therefore, MLE cannot work well with an increasing number of phases in MAP. Baum et al. \cite{Baum:AMS:1970} proposed the forward-backward algorithm for MAP in the statistical estimation of the probabilistic functions for Markov chains, which is the earliest paper for the EM algorithm. Deng et al. \cite{Deng:TS:1993} used an EM algorithm to convert the MMPP from continuous time to a discrete-time domain, obtaining an MLE containing the model parameters. Breuer \cite{Breuer:AOR:2002} designed a specification of the classical EM algorithm for the communication systems of MAP and BMAP. Roberts et al. \cite{Roberts:SPL:2006} developed a scaling forward-backward algorithm to improve Ryde\'n's EM algorithm to estimate the parameters of MMPP. Furthermore, Buchholz \cite{Buchholz:CPEMTT:2003,Buchholz:ISCIS:2004} proposed a two-step EM algorithm for fitting the PH distribution and HMM parameters based on real traffic observations. Horva\'th et al. \cite{Horvath:ICQES:2005} presented a two-step MAP fitting method. The first step fitted the PH distribution with the inter-arrival times. The second step approximated the lag correlation values. Okamura et al. \cite{Okamura:QEST:2009} improved the two-step fitting algorithm for MAP with inter-arrival time data.

\subsection{Parameter Estimation for Utilization Data}
Almost all the above studies estimated the parameters of queueing systems from observable data such as arrival times and waiting times. However, such parameter estimation approaches cannot apply to group and utilization data. Due to the greater complexity of these data types, their parameters are difficult to estimate. Okamura et al. \cite{Okamura:ITON:2009} proposed an EM algorithm for estimating the parameters of MAP for group data. Group data is defined as a group of arrival times, with each group being one bin. In each observed time interval, the number of arrival times is collected. They proposed a novel EM algorithm to calculate the expected log-likelihood function (LLF) based on the group data and then maximize the LLF to estimate the parameters.
Li et al. \cite{Li:IEEEAccess:2019} proposed the MLE approach and used the utilization data to estimate the parameters of the queueing system $M_t/M/1/K$, where the arrival process followed a nonhomogeneous Poisson process (NHPP). Since the arrival intensity of the NHPP is a continuous function of time $t$, parameter estimation is difficult. To solve the problems, they approximated the NHPP with a piecewise constant function that converted the NHPP into a series of HPPs.

This study presents an EM algorithm to estimate the parameters of a QBD queueing system with arrivals following MAP from utilization data. Compared to previous work \cite{Li:IEEEAccess:2019}, the queueing system that follows MAP arrivals is more general. A nontrivial EM algorithm is proposed to make the estimation approach work well on the queueing system. Moreover, QBD is a generalization of the birth-death process and is a powerful tool for modeling stochastic processes. The proposed EM algorithm can estimate the MAP parameters for QBD queueing systems with utilization data.

\section{Preliminaries}
\label{sec:preliminaries}
In this section, some preliminary knowledge on MAP, MMPP, and QBD process are presented, which is necessary to understand this work.

\subsection{Markovian Arrival Process}
MAP is a remarkably versatile modeling tool in point process theory. Usually, the arrival rate of MAP can be governed by CTMC. Formally, let $\MD_0$ denote the infinitesimal generator of the underlying CTMC where no arrivals occur, and $\MD_1$ denote a rate matrix that triggers a change of state of the CTMC. Then, the infinitesimal generator matrix of the CTMC is defined by $\MQ_{MAP}$ as follows:
\begin{align}
\MQ_{MAP} = 
\begin{pmatrix}
\MD_0&\MD_1\\
&\MD_0&\MD_1\\
&&\ddots&\ddots\\
&&&\MD_0&\MD_1\\
\end{pmatrix}.
\end{align}

A $m$-state MAP can be denoted as MAP($m$). Let $q_{i,j(i \neq j)}$ denote the transition rate from phase $i$ to phase $j$, where $q_{i,i} = \sum_{j=1,j \neq i}^m q_{i,j} + \sum_{j=1}^m \lambda_{i,j}$.  Then, the $\MD_0$ and $\MD_1$ matrices can be represented as follows:
\begin{align}
\MD_0 &= \begin{pmatrix}
-q_{1,1}  & q_{1,2}  & \cdots   & q_{1,m}\\
q_{2,1}   & -q_{2,2} & \cdots   & q_{2,m}\\
\vdots     & \vdots     & \ddots  & \vdots\\
q_{m,1}\ &q_{m,2}  & \cdots   & -q_{m,m}  \\
\end{pmatrix},
\label{eq:d0}
\end{align}
\begin{align}
\MD_1 &= \begin{pmatrix}
\lambda_{1,1}  & \lambda_{1,2}   & \cdots  & \lambda_{1,m} \\
\lambda_{2,1}  & \lambda_{2,2}   & \cdots  & \lambda_{2,m} \\
\vdots              & \vdots               & \ddots  & \vdots\\
\lambda_{m,1} & \lambda_{m,2}  & \cdots  & \lambda_{m,m} \\
\end{pmatrix}.
\end{align}

Let $\{N(t); t \geq 0\}$ and $\{J(t); t \geq 0\}$ represent the cumulative number of arrivals during the time interval $[0, t)$ and the phase at time $t$, respectively. In this study, $N(t)$ and $J(t)$ are called {\it level} and {\it phase}, respectively.
The infinitesimal generator of the phase process $J(t)$ can be represented by $\MD_0+\MD_1$ with the following properties:
\begin{align}
(\MD_0+\MD_1)\vone=\vzero
\end{align}
where $\vone$ and $\vzero$ are two column vectors with all elements of 1 and 0, respectively. For a MAP($m$), let $\bm{\pi}_s$ be the stationary probability vector and $\bm{\pi}_s=(\pi_{s_1},\pi_{s_2},...,\pi_{s_m})$. Then,
\begin{align}
\bm{\pi}_s(\MD_0+\MD_1)=\vzero, \quad \bm{\pi}_s \vone=1.
\end{align}
Define the fundamental arrival rate as $\tilde{\lambda}$. $\tilde{\lambda}$ can be represented by the average number of arrivals, which is given by
\begin{align}
\tilde{\lambda}=\bm{\pi}_s\MD_1\vone.
\end{align}
Define the matrix $\MP_k(t)$ whose $(i, j)$-element is given by
\begin{align}
[\MP_k(t)]_{i,j}=P(N(t)=k, J(t)=j | N(0)=0, J(0)=i).
\end{align}
Then, the differential-difference equations can be computed by
\begin{align}
\frac{d}{dt}\MP_0(t)=\MP_0(t)\MD_0,
\end{align}
\begin{align}
\frac{d}{dt}\MP_k(t)=\MP_k(t)\MD_0 + \MP_{k-1}(t)\MD_1, \quad k=1,2,3,....
\end{align}

MMPP is a doubly stochastic process where the arrival rate of a Poisson process is modulated by a CTMC, hence the name.
The MMPP can be identified as a special case of MAP, where $\MD_0$ and $\MD_1$ of the MMPP are defined by Eq. (\ref{eq:d0}) and a diagonal matrix, whose representation is
\begin{align}
\MD_1=\text{diag} \{\lambda_1,...,\lambda_m\}
\end{align}
In a word, the main difference between MAP and MMPP is whether the phase change after an arrival.

\subsection{Quasi-birth-death Process}
In general, a QBD is a special case of infinite-state CTMCs.
A continuous-time QBD Markov process can be defined by a two-dimensional process $\{N(t), J(t)\}$.
The $level$ $N(t)$ may have a finite or infinite number of states, $K$.
Assume that there is only one server in a queueing system with the finite capacity $K (\geq 1)$.
Also, the arrivals of the queueing system are served according to the FCFS discipline.
In the case of no arrival in the system, the transaction process for an arrival starts immediately.
In the case that there are less than $K$ arrivals in the system, an arrival waits to be served until all previous arrivals completing service in the system.
In the case that there are $K$ arrivals in the system, the next arrivals are refused.
Then the infinitesimal generator matrix $\MQ_{QBD}$ of the Markov process can take the $(K+1)$-by-$(K+1)$ block tridiagonal form as follows:
\begin{align}
\MQ_{QBD} = 
\left(
\begin{array}{c|ccccc}
\MB_0  & \MA_0  &           &           &            &                       \\
\hline
\MA_2  & \MA_1  & \MA_0 &           &            &                       \\
           & \ddots  & \ddots & \ddots &            &                       \\
           &            & \MA_2 & \MA_1 & \MA_0  &                       \\
           &            &           &           & \MA_2  & \MA_1 +\MA_0 \\
\end{array}\right)
\label{eq:qbd}
\end{align}
where $\MB_0$ indicates the case of state transitions at idle period, $\MA_0$ represents the state transitions with arrivals, $\MA_1$ is the state transitions with no arrival, and $\MA_2$ expresses the state transitions during the job service process.

For example, consider a specific queueing system, whose arrival process follows a MAP, and the service time follows an exponential distribution with the service rate $\mu$.
The queueing system can be indicated by the symbol $MAP/M/1/K$, and the infinitesimal generator matrix for the queueing system can be expressed by:
\begin{align}
\MQ_{QBD} = 
\begin{pmatrix}
\MD_0    & \MD_1               &                                                  \\
\mu \MI  & \MD_0-\mu \MI  & \MD_1   &                                   \\
             & \ddots               & \ddots    &                                   \\
             &                         & \mu \MI & \MD_0+\MD_1-\mu \MI \\
\end{pmatrix}
\label{eq:q_qbd}
\end{align}

\subsection{Brief Overview of the EM Algorithm}
The EM algorithm is considered as an iterative method to calculate the MLE from incomplete data \cite{Okamura:ITON:2009, Dempster:JRSSB:1977, Wu:AS:1983}.
Define $\cD$ and $\cU$ as the observable data and unobservable data, respectively.
Then, a set of model parameters $\vth$ are estimated according to the given observable data $\cD$. 
In general, the EM algorithm consists of two steps:
\begin{description}[\IEEEusemathlabelsep\IEEEsetlabelwidth{~~~~~~}]
\item[\textbf{E-step}]: Compute the expected log-likelihood function (LLF) of the complete data pair ($\cD$, $\cU$), when only the observable data $\cD$ is provided. This step can be formulated as   
\begin{align}
\text{E}_{\cU}[LLF(\vth|\cD,\cU)|\cD]
\end{align}
where $\text{E}_{\cU}$ is the expectation operator for the unobservable data $\cU$.
\item[\textbf{M-step}]: Find the parameters $\vth$ by maximizing the expected LLF. The formula is shown as follows:
\begin{align}
\label{eq:M_step}
\vth = \argmax_{\vth} \text{E}_{\cU}[LLF(\vth|\cD,\cU)|\cD]
\end{align}
\end{description}
where $E_{\cU}$ is the expectation operator for the unobservable data $\cU$.
Eq. (\ref{eq:M_step}) provides an updated formula for the parameters. That is, the estimated parameters of the M-step are used in the next E-step.
The parameters $\vth$ are updated iteratively with the two steps until the LLF or parameters converge.

\section{Parameter Esitmation for MAP}
\label{sec:estimation}
In this section, we consider the parameter estimation for a QBD process from utilization data. 
More precisely, an MLE method via the EM algorithm is considered to estimate MAP parameters.

\subsection{Utilization Data Format}
Here, we consider an EM algorithm with utilization data.
Utilization data is a kind of time-series data, which is usually defined as the time fraction of busy time over the time length of observable periods.
To estimate MAP parameters only from the utilization data, the following assumptions are given.
\begin{description}
\item[(1)] Utilization data can be monitored at every time interval.
\item[(2)] Each time interval consists of two successive periods: unobservable time interval and observable time interval.
\item[(3)] There is only one state change at most in every observable time interval. That is, a state change from busy (idle) to idle (busy).
\end{description}
The first two assumptions are reasonable according to the definition of utilization data.
Also, the third assumption is reasonable that at most only one state change can occur in every observable period when the monitoring time of the utilization data is sufficiently small.

\begin{figure*}[t]
\begin{center}
\captionsetup{justification=centering}
\includegraphics[width=0.8\hsize]{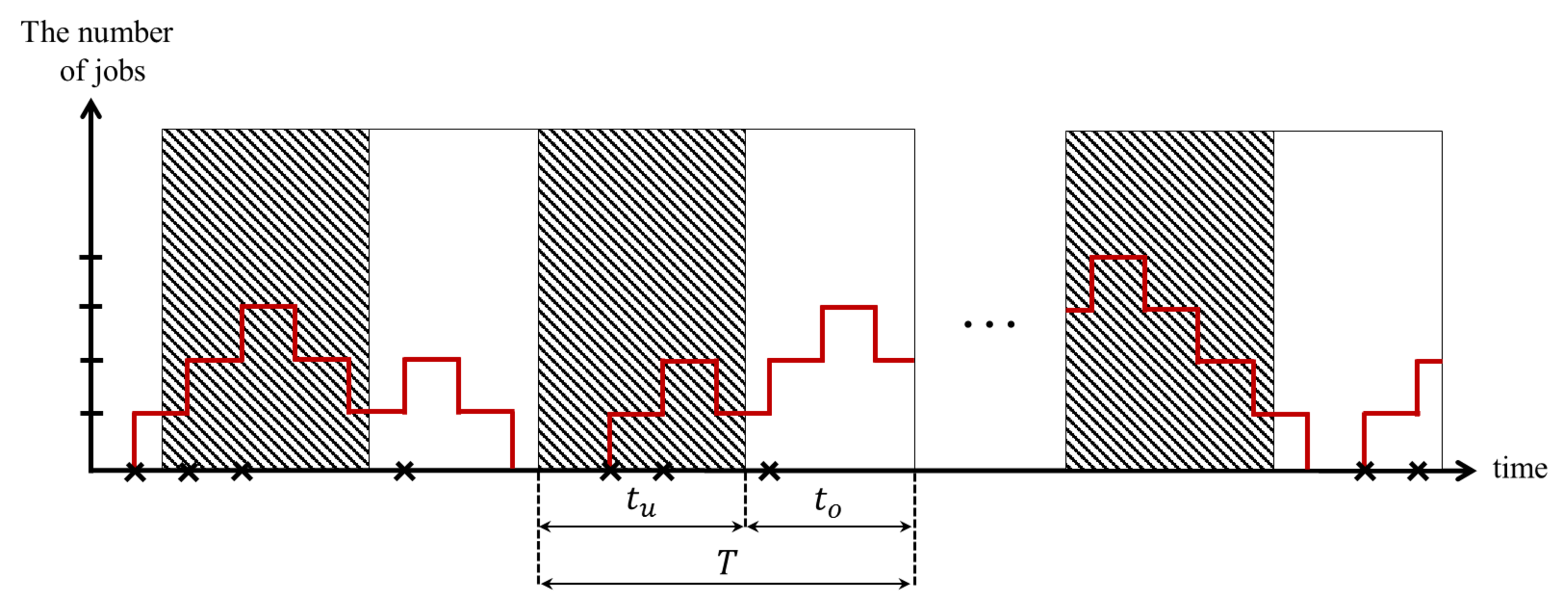}
\caption{Possible behavior of system state and observable and unobservable periods for CPU utilization.}
\label{fig:util}
\end{center}
\end{figure*}

Formally, we define a series of samples of the utilization data in a monitoring time as $\cD=(u_1, u_2, ..., u_N)$, where $u_n$ is the $n$th utilization sample of the observable period, and $0 \le u_n \le 1$.
Fig. \ref{fig:util} demonstrates an possible behavior of system state.
In the figure, let $t_u$ and $t_o$ represent the two successive unobservable time interval and observable time interval, and $t_o \ll t_u$.
Notice that the numbers of samples are counted only in the observable time interval, while the numbers of samples in the unobservable time intervals are unknown.
In other words, the sample data in the unobservable time interval are missing.
Besides, let $B_{t_o}^{(n)}$ and $I_{t_o}^{(n)}$ be the busy time and idle time for the $n$th observable time interval.
Then, the utilization is formulated by
\begin{align}
u_n=\frac{B_{t_o}^{(n)}}{B_{t_o}^{(n)}+I_{t_o}^{(n)}}
\end{align}

\subsection{A QBD Queueing System with Markovian Arrivals}
Consider a QBD queueing system, in which the jobs arrive at the system following MAP.
To formulate the MLE via the EM algorithm, the infinitesimal generator $\MQ_{QBD}$ in Eq. (\ref{eq:qbd}) are divided into four sub-matrices according to the two solid lines:
\begin{align}
&\MQ_{00}=\MB_0=\MD_0, \\
&\MQ_{01}=\left(\MA_0, \MO, ..., \MO \right)=\left(\MD_1,  \MO, ..., \MO \right),\\
&\MQ_{10}^T=\left(\MA_2,  \MO, ..., \MO \right),
\end{align}
\begin{align}
\MQ_{11} &=
\begin{pmatrix}
\MA_1  & \MA_0  &            &                       \\
\MA_2  & \MA_1  & \MA_0  &                       \\
           & \ddots  & \ddots  &                       \\
           & \MA_2  & \MA_1  & \MA_0             \\
           &            & \MA_2  & \MA_1+\MA_0  \\
\end{pmatrix}
\end{align}
where $\MO$ represents $m \times m$ zero matrix, and $\MQ_{10}^T$ indicates the matrix transpose operation of $\MQ_{10}$.
Moreover, since $\MQ$ is a $(K+1)$-by-$(K+1)$ block matrix, and each block is a $m \times m$ matrix, the dimensions of matrix $\MQ_{00}, \MQ_{01}, \MQ_{10}$ and $\MQ_{11}$ are $m \times m, m \times mK, mK \times m$ and $mK \times mK$, respectively.

Here, $\MQ_{00}$ indicates that the system is an idle state without the occurrence of state change between idle and busy, while $\MQ_{01}, \MQ_{10}$ and $\MQ_{11}$ represent that the system state changes from idle to busy, busy to idle, and busy to busy, respectively.

\subsection{M-Step Formulas}
Consider an $m$-state MAP with utilization data $\cD$.
Also, for the sake of notational convenience, we define the following several unobservable variables and observable variables.
\begin{description}
\item[$B_i$] an indicator random variable for the event that the pahse is $i$ at the initial time $t=0$.
\item[$Z_i^{[n,l]}$] cumulative sojourn time that the number of jobs is $l$ in phase $i$ during the $n$th unobservable time interval.
\item[$N_{i,j}^{[n,l]}$] the number of phase transitions from phase $i$ to phase $j$ that the number of jobs is $l$ during the $n$th unobservable time interval.
\item[$A_{i,j}^{[n,l]}$] the number of arrivals leading to phase transitions from phase $i$ to phase $j$ that the number of jobs is $l$ during the $n$th unobservable time interval.
\item[$S_{i,j}^{[n,l]}$] the number of services leading to phase transitions from phase $i$ to phase $j$ that the number of jobs is $l$ during the $n$th unobservable time interval.
\item[${\tilde Z}_i^{[n,l]}$] cumulative sojourn time that the number of jobs is $l$ in phase $i$ during the $n$th observable time interval.
\item[${\tilde N}_{i,j}^{[n,l]}$] the number of phase transitions from phase $i$ to phase $j$ that the number of jobs is $l$ during the $n$th observable time interval.
\item[${\tilde A}_{i,j}^{[n,l]}$] the number of arrivals leading to phase transitions from phase $i$ to phase $j$ that the number of jobs is $l$ during the $n$th observable time interval.
\item[${\tilde S}_{i,j}^{[n,l]}$] the number of services leading to phase transitions from phase $i$ to phase $j$ that the number of jobs is $l$ during the $n$th observable time interval.
\end{description}

Define a vector of parameters $\vth :=\{\pi_{i}, q_{i,j}, \lambda_{i,j}, \mu_{i,j}\}$, the unobservable variables $\cU :=\{B_i, Z_i^{[n,l]}, N_{i,j}^{[n,l]}, A_{i,j}^{[n,l]}, S_{i,j}^{[n,l]}\}$ and the observable variables $\cD :=\{{\tilde Z}_i^{[n,l]}, {\tilde N}_{i,j}^{[n,l]}, {\tilde A}_{i,j}^{[n,l]}, {\tilde S}_{i,j}^{[n,l]}\}$, for $i,j=1,2,...,m$ and $n=1,..,N$.
Then the MLEs of the parameters can be obtained from Eq. (\ref{eq:M_step}):
\begin{align}
\label{eq:pi}
\pi_i:\text{E}[B_i|\cD]
\end{align}
\begin{align}
\label{eq:q}
q_{i,j}:=\frac{\sum_{n=1}^{N}\text{E}\left[N_{i,j}^{[n,l]}+{\tilde N}_{i,j}^{[n,l]}|\cD\right]}{\sum_{n=1}^{N}\text{E}\left[Z_i^{[n,l]}+{\tilde Z}_i^{[n,l]}|\cD\right]},\quad i\neq j
\end{align}
\begin{align}
\label{eq:lam}
\lambda_{i,j}:=\frac{\sum_{n=1}^{N}\text{E}\left[A_{i,j}^{[n,l]}+{\tilde A}_{i,j}^{[n,l]}\right]}{\sum_{n=1}^{N}\text{E}\left[Z_i^{[n,l]}+{\tilde Z}_i^{[n,l]}|\cD\right]}
\end{align}
\begin{align}
\label{eq:mu}
\mu_{i,j}:=\frac{\sum_{n=1}^{N}\text{E}\left[S_{i,j}^{[n,l]}+{\tilde S}_{i,j}^{[n,l]}|\cD\right]}{\sum_{n=1}^{N}\text{E}\left[Z_i^{[n,l]}+{\tilde Z}_i^{[n,l]}|\cD\right]}
\end{align}

Note that the subscript of the expectation operations is omitted for the sake of simplicity.
In addition, all the expected values in the above Eq. (\ref{eq:pi})-(\ref{eq:mu}) are calculated based on the previous M-step.
Also, the EM algorithm provides an initial guess of MAP parameters at the initial step.

\subsection{E-Step Formulas}
In the E-step, the analytical forms of the expected values in Eq. (\ref{eq:pi})-(\ref{eq:mu}) are derived.
Define $T$ as the length of monitoring time interval, and $T=t_u+t_o$.
Then the cumulative time sequence for utilization data can be indicated by $s_0=0<s_1<...<s_N$, i.e., $s_n=n \times T, (n=1,2,...,N)$.
Let $\cA$ be the following event:
\begin{align}
\cA_n=\{N(s_n^+) - N(s_n^-) = a_n\}
\end{align}
where $s_n^-$ and $s_n^+$ represent the left limit and right limit, and
\begin{align}
&\{N(s_n^-)=x, N(s_n^+)=y\} \nonumber\\
&=\lim_{\Delta t \to +0}\{N(s_n-\Delta t)=x, N(s_n+\Delta t)=y\}
\end{align}
Then the forward, backward and overall events can be indicated by $\cF_n=\cA_1...\cA_n$, $\cB_n=\cA_n...\cA_N$ and $\cO=\cA_1...\cA_N$, respectively.
For simplicity, let $P(A)$ represent the probability of an indicator random variable $A$.

Define $\bm{f}(n), \tilde{\bm{f}}(n)$ and $\tilde{\bm{f}^{\prime}}(n)$ as $1 \times m(K+1)$ row vectors, who represent the probabilities (likelihoods) for the forward events in the $n$th period.
Also, define $\bm{b}(n), \tilde{\bm{b}}(n)$ and $\tilde{\bm{b}^{\prime}}(n)$ as $m(K+1) \times 1$ column vectors, who represent the probabilities (likelihoods) for the backward events in the $n$th period.
Specifically, $\bm{f}(n), \bm{b}(n)$ and $\tilde{\bm{f}}(n), \tilde{\bm{b}}(n)$represent the probability vectors at the beginning of the unobservable time interval and observable time interval, respectively.
$\tilde{\bm{f}^{\prime}}(n), \tilde{\bm{b}^{\prime}}(n)$ are the probability vectors of the system state change between busy and idle in the obserable time interval.

\begin{figure*}[t]
\begin{center}
\captionsetup{justification=centering}
\includegraphics[width=0.8\hsize]{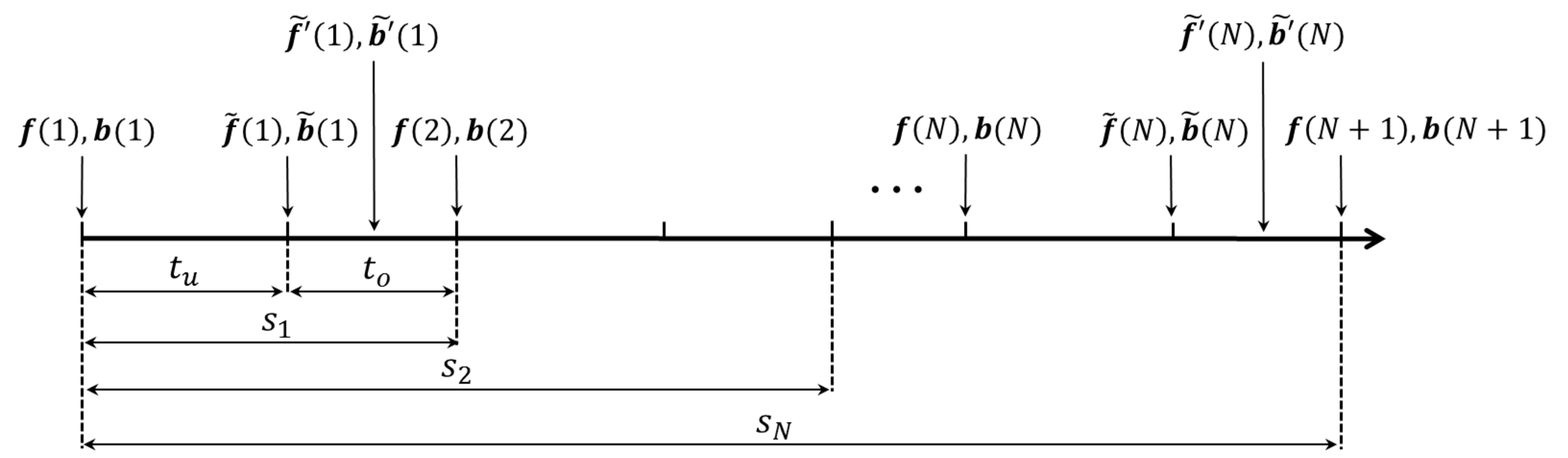}
\caption{An example of forward probabilities and backward probabilities for the system with utilization data.}
\label{fig:fb}
\end{center}
\end{figure*}

For a better understanding, fig. \ref{fig:fb} shows some examples to explain the probability vectors of forward and backward events.

Notice that each of the six vectors is partitioned by $levels$ into subvectors.
In other words, each vector has $1 \times (K+1)$ block subvector with dimension $1 \times m$.
Depending on whether the system state is idle or busy, the six vectors can be re-divided into two parts, which are demonstrated as follows:
events.
\begin{align}
\label{eq:fb01}
\begin{split}
&\bm{f}(n)=(\bm{f}_0(n), \bm{f}_1(n)), \quad \bm{b}(n)=(\bm{b}_0(n), \bm{b}_1(n)) \\
&\bm{\tilde{\bm{f}}}(n)=(\tilde{\bm{f}}_0(n), \tilde{\bm{f}}_1(n)), \quad \bm{\tilde{b}}(n)=(\tilde{\bm{b}}_0(n), \tilde{\bm{b}}_1(n)) \\
&\bm{\tilde{f}}^{\prime} (n)=(\tilde{\bm{f}}_0^{\prime} (n), \tilde{\bm{f}}_1^{\prime} (n)), \quad \bm{\tilde{b}}^{\prime} (n)=(\tilde{\bm{b}}_0^{\prime} (n), \tilde{\bm{b}}_1^{\prime} (n))
\end{split}
\end{align}
where the two elements of every vector represent the probabilities when the utilization state being idle and busy in the $n$th monitoring period, and the dimensions are $1 \times m$ and $1 \times mK$, respectively.

Consider the indicator random variable $\cO$, we can formulate the Eq. (\ref{eq:pi}) as
\begin{align}
\label{eq:pi_i}
\pi_i:=\frac{\text{E}\left[B_i\cO\right]}{P(\cO)}=\frac{\pi_i [\bm b(1)]_i}{\bm \pi \bm b(1)}
\end{align}
where $\bm{\pi}$ be the initial probability vector, and $\bm{\pi}=(\pi_1,\pi_2,...,\pi_m)$ with $\sum_{i=1}^{m}\pi_i=1$.

Next, according to the Eq. (\ref{eq:q}), we consider the expected values $\text{E}\left[Z_i^{[n,l]}|\cD\right], \text{E}\left[\tilde Z_i^{[n,l]}|\cD\right], \text{E}\left[N_{i,j}^{[n,l]}|\cD\right]$ and $\text{E}\left[\tilde N_{i,j}^{[n,l]}|\cD\right]$, respectively. For the calculation of $\text{E}\left[Z_i^{[n,l]}|\cD\right]$, since $\text{E}\left[Z_i^{[n,l]}|\cD\right]= \text{E}\left[Z_i^{[n,l]}\cO\right]/P(\cO)$, the subsequent analysis treats only $\text{E}\left[Z_i^{[n,l]}\cO\right]$, which can be formulated by
\begin{align}
&\text{E}\left[Z_i^{[n,l]}\cO\right]=\nonumber\\
&\int_{0}^{t_u} \left[\bm{f}(n)\exp\left(\MQ s\right)\right]_{(l,i)}\left[\exp\left(\MQ(t_u-s)\right)\tilde{\bm{b}}(n)\right]_{(l,i)}ds 
\end{align}
Note that $[\cdot]_{(l,i)}$ indicates the $i$th element of the $l$th block in the block vector $[\cdot]$. 

Similarly, $\text{E}\left[\tilde Z_i^{[n,l]}|\cD\right]$ can be derived by
\begin{align}
\text{E}\left[\tilde{Z}_i^{[n,l]}|\cD\right]=\frac{\text{E}\left[\tilde{Z}_i^{[n,l]}\cO\right]}{P(\cO)}
\end{align}
For $\text{E}\left[\tilde{Z}_i^{[n,l]}\cO\right]$, the calculations are given by the two equations as follows according to the number of jobs $l$.

if $l=0$:
\begin{align}
&\text{E}\left[\tilde{Z}_i^{[n,l]}\cO\right]=
\int_{0}^{(1-u_n)t_o} \left[\tilde{\bm f}_0(n)\exp\left(\MQ_{00} s\right)\right]_i\nonumber\\
&\times \left[\exp\left(\MQ_{00}\left((1-u_n)t_o-s\right)\right)\MQ_{01}\tilde{\bm b}_1^{\prime}(n)\right]_i ds \nonumber\\
&+ \int_{0}^{(1-u_n)t_o}\left[\tilde{\bm f}_1^{\prime}(n)\MQ_{10}\exp\left(\MQ_{00} s\right)\right]_i \nonumber\\
&\times \left[\exp\left(\MQ_{00}\left((1-u_n)t_o-s\right)\right){\bm b}_0(n+1)\right]_i ds
\end{align}
where $[\cdot]_i$ indicates the $i$th element of vector $[\cdot]$.

if $l>0$:
\begin{align}
&\text{E}\left[\tilde{Z}_i^{[n,l]}\cO\right]=\int_{0}^{u_n t_o} \left[\tilde{\bm f}_1(n)\exp\left(\MQ_{11} s\right)\right]_{(l,i)}\nonumber\\
&\times \left[\exp\left(\MQ_{11}(u_n t_o-s)\right)\MQ_{10}\tilde{\bm b}_0^{\prime}(n)\right]_{(l,i)}ds \nonumber\\
&+ \int_{0}^{u_n t_o}\left[\tilde{\bm f}_0^{\prime}(n)\MQ_{01}\exp\left(\MQ_{11} s\right)\right]_{(l,i)}\nonumber\\
&\times \left[\exp\left(\MQ_{11}(u_n t_o-s)\right){\bm b}_1(n+1)\right]_{(l,i)}ds 
\end{align}

Similarly, $\text{E}\left[N_{i,j}^{[n,l]}|\cD\right]$ can be derived by
\begin{align}
\text{E}\left[N_{i,j}^{[n,l]}|\cD\right]=\frac{\text{E}\left[N_{i,j}^{[n,l]}\cO\right]}{P(\cO)}
\end{align}
\begin{align}
&\text{E}\left[N_{i,j}^{[n,l]}\cO\right]=
\int_{0}^{t_u} \left[\bm{f}(n) \exp\left(\MQ s\right)\right]_{(l,i)}[\MB_0]_{i,j}\nonumber\\
&\times \left[\exp\left(\MQ(t_u-s)\right)\tilde{\bm b}(n)\right]_{(l,j)}ds
\end{align}
where $[\cdot]_{i,j}$ represents the $(i,j)$th element of the matrix $[\cdot]$.

In addition, $\text{E}\left[\tilde N_{i,j}^{[n,l]}\right]$ can be derived by
\begin{align}
\text{E}\left[\tilde{N}_{i,j}^{[n,l]}|\cD\right]=\frac{\text{E}\left[\tilde{N}_{i,j}^{[n,l]}\cO\right]}{P(\cO)}
\end{align}
According to the number of jobs $l$, the computation of $\text{E}\left[\tilde{N}_{i,j}^{[n,l]}\cO\right]$ are as follows.

if $l=0$:
\begin{align}
&\text{E}\left[\tilde{N}_{i,j}^{[n,l]}\cO\right]=
\int_{0}^{(1-u_n)t_o} \left[\tilde{\bm f}_0(n) \exp\left(\MQ_{00} s\right)\right]_i [\MB_0]_{i,j}\nonumber\\
&\times \left[\exp\left(\MQ_{00}\left((1-u_n)t_o-s\right)\right)\MQ_{01}\tilde{\bm b}_1^{\prime}(n)\right]_j ds \nonumber\\
&+ \int_{0}^{(1-u_n)t_o}\left[\tilde{\bm f}_1^{\prime}(n)\MQ_{10}\exp\left(\MQ_{00} s\right)\right]_i [\MB_0]_{i,j}\nonumber\\
&\times \left[\exp\left(\MQ_{00}\left((1-u_n)t_o-s\right)\right){b}_0(n+1)\right]_j ds
\end{align}

if $l>0$:
\begin{align}
&\text{E}\left[\tilde{N}_{i,j}^{[n,l]}\cO\right]=
\int_{0}^{u_n t_o} \left[\tilde{\bm f}_1(n) \exp\left(\MQ_{11} s\right)\right]_{(l,i)}[\MB_0]_{i,j}\nonumber\\
&\times \left[\exp\left(\MQ_{11}(u_n t_o-s)\right)\MQ_{10}\tilde{\bm b}_0^{\prime}(n)\right]_{(l,j)}ds \nonumber\\
&+ \int_{0}^{u_n t_o}\left[\tilde{\bm f}_0^{\prime}(n)\MQ_{01}\exp\left(\MQ_{11} s\right)\right]_{(l,i)}[\MB_0]_{i,j}\nonumber\\
&\times \left[\exp\left(\MQ_{11}(u_n t_o-s)\right){\bm b}_1(n+1)\right]_{(l,j)}ds 
\end{align}

Next, according to the Eq. (\ref{eq:lam}), we consider the expected values $\text{E}\left[A_{i,j}^{[n,l]}|\cD\right]$ and $\text{E}\left[\tilde A_{i,j}^{[n,l]}|\cD\right]$, respectively. 
$\text{E}\left[A_{i,j}^{[n,l]}|\cD\right]$ can be derived by
\begin{align}
\text{E}\left[A_{i,j}^{[n,l]}|\cD\right]=\frac{\text{E}\left[A_{i,j}^{[n,l]}\cO\right]}{P(\cO)}
\end{align}
\begin{align}
&\text{E}\left[A_{i,j}^{[n,l]}\cO\right]=
\int_{0}^{t_u} \left[\bm{f}(n) \exp\left(\MQ s\right)\right]_{(l,i)}[\MA_0]_{i,j}\nonumber\\
&\times \left[\exp\left(\MQ(t_u-s)\right)\tilde{\bm b}(n)\right]_{(l+1,j)}ds
\end{align}

Similarly, $\text{E}\left[\tilde A_{i,j}^{[n,l]}\right]$ is derived by 
\begin{align}
\text{E}\left[\tilde{A}_{i,j}^{[n,l]}|\cD\right]=\frac{\text{E}\left[\tilde{A}_{i,j}^{[n,l]}\cO\right]}{P(\cO)}
\end{align}
 
 if $l=0$:
\begin{align}
&\text{E}\left[\tilde{A}_{i,j}^{[n,l]}\cO\right]=
\int_{0}^{(1-u_n)t_o} \left[\tilde{\bm f}_0(n) \exp\left(\MQ_{00} s\right)\right]_i [\MA_0]_{i,j}\nonumber\\
&\times \left[\exp\left(\MQ_{00}\left((1-u_n)t_o-s\right)\right)\MQ_{01}\tilde{\bm b}_1^{\prime}(n)\right]_j ds
\end{align}
 
 if $l>0$:
\begin{align}
&\text{E}\left[\tilde{A}_{i,j}^{[n,l]}\cO\right]=
\int_{0}^{u_n t_o} \left[\tilde{\bm f}_1(n)\exp\left(\MQ_{11} s\right)\right]_{(l,i)}[\MA_0]_{i,j}\nonumber\\
&\times \left[\exp\left(\MQ_{11}(u_n t_o-s)\right)\MQ_{10}\tilde{b}_0^{\prime}(n)\right]_{(l+1,j)}ds \nonumber\\
&+ \int_{0}^{u_n t_o}\left[\tilde{\bm f}_0^{\prime}(n)\MQ_{01}\exp\left(\MQ_{11} s\right)\right]_{(l,i)}[\MA_0]_{i,j}\nonumber\\
&\times \left[\exp\left(\MQ_{11}(u_n t_o-s)\right){b}_1(n+1)\right]_{(l+1,j)}ds
\end{align}

Finally, we calculate $\text{E}\left[S_{i,j}^{[n,l]}|\cD\right]$ and $\text{E}\left[\tilde{S}_{i,j}^{[n,l]}|\cD\right]$ according to the Eq. (\ref{eq:mu}).
$\text{E}\left[S_{i,j}^{[n,l]}|\cD\right]$ is derived by:
\begin{align}
\text{E}\left[S_{i,j}^{[n,l]}|\cD\right]=\frac{\text{E}\left[S_{i,j}^{[n,l]}\cO\right]}{P(\cO)}
\end{align}
Then, $\text{E}\left[S_{i,j}^{[n,l]}\cO\right]$ can be expressed by
\begin{align}
&\text{E}\left[S_{i,j}^{[n,l]}\cO\right]=
\int_{0}^{t_u} \left[\bm{f}(n)\exp\left(\MQ s\right)\right]_{(l,i)}[\MA_2]_{i,j}\nonumber\\
&\times \left[\exp\left(\MQ(t_u-s)\right)\tilde{\bm b}(n)\right]_{(l-1,j)}ds
\end{align}

Similarly, $\text{E}\left[\tilde{S}_{i,j}^{[n,l]}|\cD\right]$ derived by
\begin{align}
\text{E}\left[\tilde{S}_{i,j}^{[n,l]}|\cD\right]=\frac{\text{E}\left[\tilde{S}_{i,j}^{[n,l]}\cO\right]}{P(\cO)}
\end{align}
According to the number of jobs $l$, the equation $\text{E}\left[\tilde{S}_{i,j}^{[n,l]}\cO\right]$ can be divided into two parts according to $l$:

if $l=1$:
\begin{align}
&\text{E}\left[\tilde{S}_{i,j}^{[n,l]}\cO\right]=
\int_{0}^{(1-u_n)t_o} \left[\tilde{\bm f}_1(n)\exp\left(\MQ_{11} s\right)\right]_{(l,i)} [\MA_2]_{i,j}\nonumber\\
&\times \left[\exp\left(\MQ_{11}\left((1-u_n)t_o-s\right)\right)\MQ_{10}\tilde{\bm b}_0^{\prime}(n)\right]_{(l-1,j)} ds
\end{align}
 
if $l>1$:
\begin{align}
\label{eq:S_ij}
&\text{E}\left[\tilde{S}_{i,j}^{[n,l]}\cO\right]=
\int_{0}^{u_n t_o} \left[\tilde{\bm f}_1(n)\exp\left(\MQ_{11} s\right)\right]_{(l,i)}[\MA_2]_{i,j}\nonumber\\
&\times \left[\exp\left(\MQ_{11}(u_n t_o-s)\right)\MQ_{10}\tilde{\bm b}_0^{\prime}(n)\right]_{(l-1,j)}ds \nonumber\\
&+ \int_{0}^{u_n t_o}\left[\tilde{\bm f}_0^{\prime}(n)\MQ_{01}\exp\left(\MQ_{11} s\right)\right]_{(l,i)}[\MA_2]_{i,j}\nonumber\\
&\times \left[\exp\left(\MQ_{11}(u_n t_o-s)\right){\bm b}_1(n+1)\right]_{(l-1,i)}ds
\end{align}

\subsection{Computation of E-step Formulas}
The E-step of MAP estimation with utilization data requires the computation of the probabilities of the forward and backward events and their convolutions.

According to fig. \ref{fig:fb}, we first derive the likelihoods for forward events in the $n$th period.
Since the state transitions in unobservable intervals can not be observed, $\tilde{\bm{f}}(n)$ is computed by
\begin{align}
\label{eq:hatf}
\tilde{\bm{f}}(n)=\bm{f}(n)\exp\left(\MQ t_o\right)
\end{align}
In the $n$th observable interval, according to the idle state and busy state of the system, the computation of the likelihood $\tilde{\bm{f}}^{\prime}(n)$ are divided into two parts, which are expressed as:  
\begin{align}
\tilde{\bm f}_0^{\prime} (n)=\tilde{\bm f}_0(n) \exp\left(\MQ_{00}(1-u_n) t_o\right)
\end{align}
\begin{align}
\tilde{\bm f}_1^{\prime} (n)=\tilde{\bm f}_1(n) \exp\left(\MQ_{11}u_n t_o\right)
\end{align}
Also, the instantaneous transition probability $\bm{f}(n+1)$ can be calculated based on $\tilde{\bm f}^{\prime} (n)$. 
According to Eq. \ref{eq:fb01}, $\bm{f}_0(n+1)$ and  $\bm{f}_1(n+1)$ are shwon as follows:

if $u_n=0$:
\begin{align}
\bm{f}_0(n+1)=\tilde{\bm f}_0^{\prime}(n)
\end{align}
\begin{align}
\bm{f}_1(n+1)=\vzero_{1\times m(K+1)}
\end{align}

if $0<u_n<1$:
\begin{align}
\bm{f}_0(n+1)=\tilde{\bm f}_1^{\prime}(n) \MQ_{10}\exp\left(\MQ_{00}(1-u_n) t_o\right)
\end{align}
\begin{align}
\bm{f}_1(n+1)=\tilde{\bm f}_0^{\prime}(n) \MQ_{01} \exp\left(\MQ_{11} u_n t_o\right)
\end{align}

if $u_n=1$:
\begin{align}
\bm{f}_0(n+1)=\vzero_{1\times m(K+1)}
\end{align}
\begin{align}
\label{eq:f1}
\bm{f}_1(n+1)=\tilde{\bm f}_1^{\prime}(n)
\end{align}
Then, the convolution of the $n$th forward likelihood $\bm{f}(n)$ can be calculated according to the Eq. (\ref{eq:hatf})-(\ref{eq:f1}).

Similarly, the computations of backward likelihoods in $n$th period can be expressed as follows.
\begin{align}
\label{eq:b}
\bm{b}(n)=\exp\left(\MQ t_o\right)\tilde{\bm b}(n)
\end{align}
\begin{align}
\tilde{\bm b}_0^{\prime} (n)=\exp\left(\MQ_{00}(1-u_n) t_o\right) \bm{b}_0(n+1)
\end{align}
\begin{align}
\tilde{\bm b}_1^{\prime} (n)=\exp\left(\MQ_{11}u_n t_o\right) \bm{b}_1(n+1)
\end{align}

if $0<u_n<1$:
\begin{align}
\tilde{\bm b}_0(n)=\exp\left(\MQ_{00}(1-u_n) t_o\right)\MQ_{01}\tilde{\bm b}_1^{\prime}(n)
\end{align}
\begin{align}
\tilde{\bm b}_1(n)=\exp\left(\MQ_{11}u_n t_o\right)\MQ_{10}\tilde{\bm b}_0^{\prime}(n)
\end{align}

if $u_n=0$:
\begin{align}
\tilde{\bm b}_0(n)=\tilde{\bm b}_0^{\prime}(n)
\end{align}
\begin{align}
\tilde{\bm b}_1(n)=\vzero_{m(K+1)\times 1}
\end{align}

if $u_n=1$:
\begin{align}
\tilde{\bm b}_0(n)=\vzero_{m(K+1)\times 1}
\end{align}
\begin{align}
\label{eq:hatb}
\tilde{\bm b}_1(n)=\tilde{\bm b}_1^{\prime}(n)
\end{align}
Then, $\bm{b}(1)$ can be derived according to the convolution equations above of Eq.(\ref{eq:b})-(\ref{eq:hatb}).

In general, the computations of $\bm{f}(n)$ and $\bm{b}(1)$ take too much cost with the number of phases $m$ and the capacity $K$ increasing.
To reduce the computation, $\bm{f}(n)$ and $\bm{b}(1)$ can be computed by using uniformization technique \cite{Okamura:TAAODAIUS:2008}.
Let $q$ be a positive constant, and $q$ is larger than the maximum value of the absolute diagonal element of $\MD_0$.
Then, the equation for the uniformization is given by
\begin{align}
\exp \left(\MD_0t\right)=\sum_{z=0}^{\infty} e^{-qt}\frac{(qt)^z}{z!}(\MI+\MD_0/q)
\end{align}
where $\MI$ is the $m\times m$ identity matrix.
Since there exists the sum operation of the above equation, whose upper limit is $\infty$, we usually truncate it by a certain point in the practical computation.
In general, a simple and effective way to truncate the upper limit is based on Poisson probability mass function.

Now, we summarize the EM algorithm to estimate the parameters of the MAP with utilization data.
\begin{algorithm}[H]
\caption{EM Algorithm with Utilization Data}
\label{alg:em}
\begin{algorithmic}[1]
\item[Step 1:] Determine the initial parameters $\vth:=$$\{\bm \pi_i, q_{i,j},$ $\lambda_{i,j},$ $\mu_{i,j}\}$
\item[Step 2:] Compute $\bm{f}(n)$, $\bm{b}(n)$, $\tilde{\bm{f}}(n)$, $\tilde{\bm{b}}(n)$, $\tilde{\bm{f}^{\prime}}(n)$, $\tilde{\bm{b}^{\prime}}(n)$ according to the Eq.(\ref{eq:b})-(\ref{eq:hatb}).
\item[Step 3:] Calcluate the expected values of $\text{E}\left[Z_i^{[n,l]}|\cD\right]$, $\text{E}\left[\tilde Z_i^{[n,l]}|\cD\right]$, $\text{E}\left[N_{i,j}^{[n,l]}|\cD\right]$, $\text{E}\left[\tilde N_{i,j}^{[n,l]}|\cD\right]$, $\text{E}\left[A_{i,j}^{[n,l]}|\cD\right]$, $\text{E}\left[\tilde {A}_{i,j}^{[n,l]}|\cD\right]$, $\text{E}\left[S_{i,j}^{[n,l]}|\cD\right]$, and $\text{E}\left[{\tilde S}_{i,j}^{[n,l]}|\cD\right]$ according to Eq. (\ref{eq:pi_i})-(\ref{eq:S_ij}), respectively.
\item[Step 4:] Update the parameters according to Eq. (\ref{eq:pi})-(\ref{eq:mu}).
\item[Step 5:] Stop the algorithm if the converge condition is satisfied. Otherwise, go to Step 1.
\end{algorithmic}
\end{algorithm}

In step 5, we need to determine the converge condition to out the iteration process.
A simple method of the condition to stop the iteration is based on the difference between the two successive likelihoods or the relative difference between the two successive likelihoods.
The coverage condition is formulated as
\begin{align}
|\text{LLF}(\vth^\prime)-\text{LLF}(\vth)| < \epsilon_a,~~\frac{|\text{LLF}(\vth^\prime)-\text{LLF}(\vth)|}{|\text{LLF}(\vth)|}< \epsilon_r
\end{align}
where $\vth^\prime$ and $\vth$ represent the parameters at the current iteration and the previous iteration.
Notice that in this paper, we use $P(\cO)=\bm \pi \bm{b}(1)$ as the LLF.

\subsection{Optimal Model Selection}
Consider a $m$-state MAP.
The value $m$ determines the number of phases of a MAP.
In general, the accuracy of the model improves with the number of phases $m$ increases.
However, a large $m$ sometimes may cause the overfitting problem, so that the accuracy of fitting gets worse.
In our previous work \cite{Li:IEEEAccess:2019}, we used AIC \cite{Sakamoto:AIC:1986} to select the optimal model for $NHPP/M/1/K$ queueing systems.
And the experiments were verified that AIC can work well in selecting optimal queueing models.
In this paper, we also use the AIC, which can quantify the goodness of fit of the model.
The formula below definites the AIC:
\begin{align}
\text{AIC}=-2\text{LLF}+2\left(\#~\text{of free parameters}\right)
\end{align}

In the case of general MAP with $m$ phases, the number of free parameters in $\bm\pi, \MD_0$ and $\MD_1$ is $2m^2-1$.
On the other hand, there are $m^2$ free parameters in an MMPP.
Then, the optimal number of phases can be computed by determining the smallest value of the AIC.







\section{Conclusion}
\label{sec:conclusion}
This paper investigated the parameter estimation problem for MAP-driven QBD queueing systems using utilization data. Unlike conventional approaches that rely on inter-arrival times, waiting times, or queue-length observations, utilization data only records the proportion of busy time within each observable monitoring interval. As a result, key information such as arrival epochs, service completions, phase transitions, and queue-length trajectories is partially or completely unobserved. To address this incomplete-data setting, we formulated the queueing system as a QBD process and developed an EM algorithm for maximum likelihood estimation of the MAP and service parameters.

The proposed method explicitly separates observable and unobservable intervals and derives the expected sufficient statistics required in the E-step, including cumulative sojourn times, phase transitions, arrivals, and service completions. These quantities are then used in the M-step to update the model parameters iteratively. In addition, AIC-based model selection is incorporated to determine the number of MAP phases, providing a practical way to balance fitting accuracy and model complexity.

The main significance of this study is that it enables MAP parameter estimation from utilization data, which is much easier to obtain in practical computer systems than detailed event-level observations. By combining MAP, QBD modeling, and EM-based inference, the proposed framework extends queueing parameter estimation to a more realistic monitoring scenario. Future work will include more extensive validation with real system traces, robustness analysis under different monitoring resolutions, and extensions to more general service-time distributions and multi-server queueing systems.

\end{document}